\documentclass[11pt,fleqn]{article}
\usepackage{amsmath,amssymb,amsthm,epsfig,graphics}

\def\R{\mathbb R}
\def\Z{\mathbb Z}

\newtheorem{theorem}{Theorem}

\newtheorem*{lemma*}{Lemma}

\newtheorem*{conjecture*}{Conjecture}

{\theoremstyle{definition}
\newtheorem{definition}{Definition}
\newtheorem{example}{Example}

\newtheorem*{note*}{Note}
}

\allowdisplaybreaks

\begin{document}

\begin{center}
\LARGE \bf
New families of cryptographic systems
\end{center}

\begin{center} \bf
Maryna NESTERENKO~$^{\dag\S}$, Jiri PATERA~$^\dag$ and Dmytro ZHAVROTSKYI~$^\ddag$
\end{center}

\noindent $^\dag$~Centre de Recherches Math\'ematiques , Universit\'e de Montr\'eal,
6128, succursale Centre-ville,\\
$\phantom{^\dag}$~Montr\'eal, QC, H3C3J7, Canada\\
$\phantom{^\dag}$~E-mail: patera@crm.umontreal.ca

\noindent $^\S$~Institute of Mathematics of NAS of Ukraine, 3
Tereshchenkivs'ka Str., Kyiv-4, 01601 Ukraine\\
$\phantom{^\S}$~E-mail: maryna@imath.kiev.ua

\noindent $^\ddag$~Interzvyazok, 15 Novokostyantynivska str.,
Kyiv, 04080, Ukraine\\
$\phantom{^\ddag}$~E-mail: dima\_zh@isv.com.ua

\begin{abstract}
\noindent
A symmetric encryption method based on
properties of quasicrystals is proposed.
The advantages of the cipher are strict aperiodicity
and everywhere discontinuous property as well as
the speed of computation, simplicity of implementation and
a straightforward possibility of
extending the method to encryption of higher dimensional data.
\end{abstract}

\section{Introduction}\label{Introduction}
 Here we address the `classical' cryptographic problem, i.e.
protection of transmitted and stored data from divulgence and
distortion. Such problem frequently arise when data creation or/and
data usage are disjoint in time or/and space. There exist many
approaches to this problem in modern cryptography, which can be
divided into three main groups in accordance with their key
primitives, i.e., unkeyed, symmetryc-key and public-key encodings,
see~\cite{Diffie-Hellman1976, Goldreich2001,
Menezes-Oorschot-Vanstone2001}.

In the present work we propose a new generic encoding procedure
based on the aperiodic point sets, called quasicrystals in the
physics literature and model sets in the mathematics
literature~\cite{CMP, Gazeau-Masakova-Pelantova2006}.
Generally speaking, such an algorithm is a symmetric stream cipher endowed
with strict aperiodicity (no periodic subsets) and everywhere
discontinuous properties.

The paper starts by preliminaries and general encoding idea,
which is followed by the precise statement
of pertinent mathematical ingredients, Sections II and III.
In Section~\ref{sec_image_coding} applications
of the proposed cipher to encoding of bitmap pictures are presented and discussed.
The advantages of the approach are in the speed of computation
and a straightforward possibility of
extending the method to encryption of 3 and higher dimensional data.
Possible range of private keys and some other applications of quasicrystals
are discussed in Conclusions.

\section{Preliminaries and encryption procedure}\label{sec_preliminaries}
We call cut-and-project quasicrystal a discrete deterministic aperiodic
point set $\Lambda$, in a finite-dimensional real Euclidean space.
In many ways such quasicrystals resemble lattices in all but the translation  symmetry.
Sometimes they are even called
aperiodic lattices or aperiodic crystals.
In this work we make use
of two remarkable properties of the quasicrystals:
(i) no periodic subsets of any kind are contained in
a quasicrystal, and
(ii) the discontinuity of the `star map' between a quasicrystal $\Lambda$ and its `acceptance window' $\Omega$.

\begin{figure}[ht]
\centering
\includegraphics[scale=1]{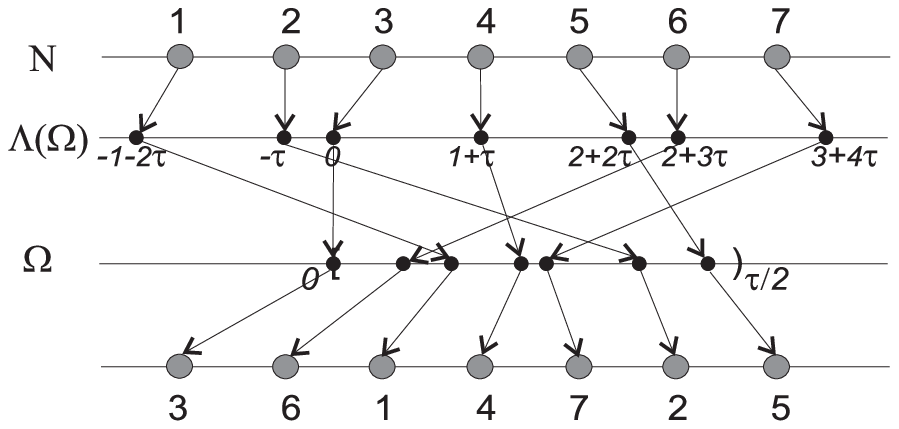}
\caption{Scheme of the simplest quasicrystal-based cipher}\label{fig_shema}
\end{figure}

General idea of the proposed approach is to take given digital data,
set a one-to-one correspondence between information bites and integer numbers $N$, and then
to map points of $N$ on a quasicrystal fragment $\Lambda$
(see the first two lines of~Fig.~\ref{fig_shema}).
At this stage the data is mapped to the quasicrystal points.
It is followed by the application of the star map to the quasicrystal points
(see the second and the third lines of~Fig.~\ref{fig_shema}), which is a crucial step of the encryption.
As a result, we get the quasicrystal points,
carrying our data, `tossed up' in the acceptance window $\Omega$.
It remains to map the content of $\Omega$
on a desirable lattice for the output
(see the last two lines of Fig.~\ref{fig_shema}).

Decryption proceeds in the opposite direction.

The set of private keys in our encryption method is infinite
and is discussed in Conclusions. In addition, one may choose to use any common substitution scheme.

Generation of quasicrystal points is very flexible.
It can either be done as a single stream or split into several blocks. Moreover, calculation of the points of each block is easily amenable to parallel computing.

Using quasicrystals, all calculations can be carried out in integers.
That guarantees absolute accuracy of coding and decoding procedures, assures the stability of the algorithm.

\section{Mathematical ingredients}\label{sec_mathematics}
\subsection{Theoretical background}
Theoretical background of the work can be found in~\cite{CMP,
Gazeau-Masakova-Pelantova2006, Guimond-Masakova-Pelantova2003,
Masakova-Patera-Pelantova1998a, MPP, Moody-Patera1993}.

\begin{figure}[ht]
\centerline{
\includegraphics[scale=0.6]{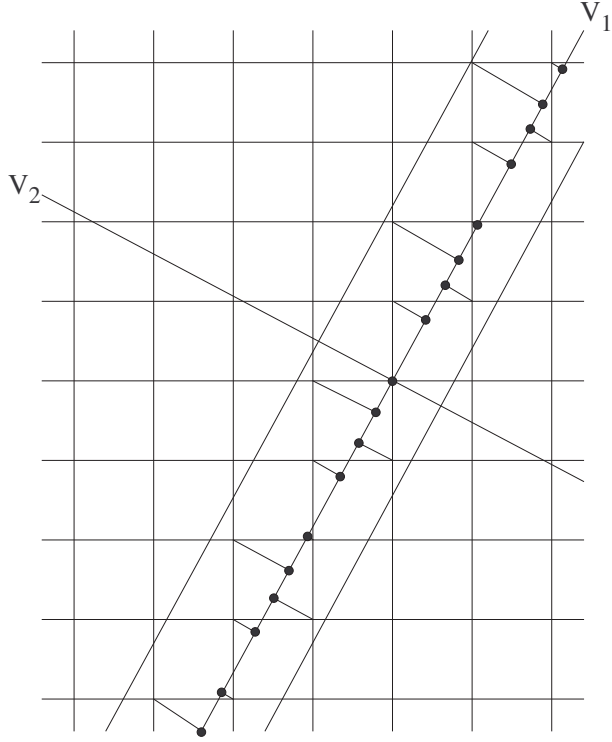}}
\caption{Construction of one-dimensional quasicrystal}\label{fig_cut-project}
\end{figure}

Our aim is to consider two-dimensional data.
We start with a choice of two straight lines $V_1\colon y=\tau x$ and $V_2\colon y=\tau' x$
on a~square lattice $\Z_2$ (see Fig.~\ref{fig_cut-project}),
where $\tau$ and $\tau'$ are two irrational numbers to be specified later. The simplest choice is
$\tau=\frac 12(1+\sqrt 5)$ and $\tau'=\frac 12(1-\sqrt 5)$.
One of the lines, say $V_1$, plays the role of the quasicrystal space onto which a `cut' of the  lattice $\Z_2$ is projected.
In the other direction $V_2$, we choose a finite interval $\Omega$.
Point of the lattice $\Z_2$ becomes a~point of the quasicrystal, after projection on $V_1$, provided its projection on $V_2$ falls within $\Omega$, i.e. it is within the `cut'.

We denote by $\Z[\tau]$ all the real numbers of the form $(a+b\tau)$, with integers $a$ and $b$.

The \textit{star map} between a quasicrystal point $(a+b\tau)$ and the point $(a+b\tau')$ of $\Omega$ is given by
\begin{gather*}
\begin{tabular}{rcl}
$\star\ \colon\quad \Z[\tau]$ &$\longrightarrow$&$\Z[\tau']$,\\
$(a+b\tau)$&$\mapsto$& $(a+b\tau')$,\qquad $a,b\in\Z$.
\end{tabular}
\end{gather*}

There is an unlimited number of choices of the irrational pairs
to use in our construction. For example, one may choose as $\tau$ and $\tau'$
the solutions of the two infinite series of the quadratic equations $x^2=mx+1,\ m=1,2,\ldots$
and $x^2=mx-1,\ m=3,4,\ldots$.

The one-dimensional \textit{quasicrystal} or \textit{cut and project set} $\Lambda_{\tau}(\Omega)$ is described as follows,

\begin{definition}
Let $\Omega$ be a finite interval and $\tau$, $\tau'$
be irrational numbers, then
\begin{gather*}
\Lambda_{\tau}(\Omega)=\{a+b\tau\mid a,b\in\Z,\ a+b\tau'\in\Omega\},
\end{gather*}
where $\Omega$ is called the \textit{acceptance window} of $\Lambda_{\tau}(\Omega)$.
\end{definition}

Let us underline, that $\Lambda_{\tau}(\Omega)$
defined here is an infinite, uniformly dense, and uniformly discrete point set, while its window $\Omega$ is a finite interval densely covered by the star map of the points of $\Lambda_{\tau}(\Omega)$.

From the definition of $\Lambda_{\tau}(\Omega)$ a number of properties of $\Lambda_{\tau}(\Omega)$
can be shown~\cite{Gazeau-Masakova-Pelantova2006, Moody-Patera1993}.
In particular, the aperiodicity, discontinuity of the star map and the existence of only two or three different distances
between adjacent points of $\Lambda_{\tau}(\Omega)$.
More precisely, we recall the following theorems.

\begin{theorem}
Let $\Omega$ be a semi-closed interval. For every $\Lambda_{\tau}(\Omega)$ there
exist positive numbers $\Delta_1,\Delta_2\in \Z[\tau]$,
such that the distances between adjacent points in $\Lambda_{\tau}(\Omega)$
take values in $\{\Delta_1,\Delta_2,\Delta_1+\Delta_2\}$.
The distances depend only on $\tau,\tau'$ and length of the acceptance window.
\end{theorem}

More can be said about the distances between adjacent points, namely
\begin{gather*}
\{x_{n+1}-x_n|n\in\Z\}=
\left\{
\begin{array}{l}
\{\Delta_1,\ \Delta_2,\ \Delta_1+\Delta_2\}, \\
\text{when}\quad \Delta_1^\star-\Delta_2^\star>|\Omega|
\\[1 ex]
\{\Delta_1,\ \Delta_2\},\\
\text{when}\quad \Delta_1^\star-\Delta_2^\star=|\Omega|.
\end{array}
\right.
\end{gather*}

One has a general scaling property of quasicrystals:
\begin{theorem}
For every irrational numbers $\tau,\tau'$, $\tau\ne\tau'$ and a bounded interval $\Omega$,
there exist $\tilde\tau'\in(-1,0)$, $\tilde\tau>0$, $s\in\R$ and
$\tilde\Omega$ (satisfying $\max\{1+\tilde\tau',-\tilde\tau'\}<|\tilde\Omega|\le 1$),
such that
\begin{gather*}
\Lambda_{\tau}(\Omega)=s\Lambda_{\tilde\tau}(\tilde\Omega).
\end{gather*}

Moreover, the distances between adjacent points in $\Lambda_{\tilde\tau}(\tilde\Omega)$ are equal to
$\tilde\tau$, $1+\tilde\tau$, $1+2\tilde\tau$, when $|\tilde\Omega|\ne 1$ and to
$\tilde\tau$, $1+\tilde\tau$, when $|\tilde\Omega|=1$.
\end{theorem}

Finally we underline that the 1-dimensional quasicrystal $\Lambda_{\tau}(\Omega)$ is an infinite discrete
aperiodic set of points uniquely determined by the choice of $\tau$, $\tau'$ and  of
acceptance window $\Omega=[c,c+d)$, where $c$ and $d$ are any real numbers.
Due to the theorem~2, without loss of generality, we can consider
$\tau'\in(-1,0)$, $\tau>0$, $c=0$ and
$d\in(\max\{1+\tau',-\tau'\},1]$.

In higher dimensions, it is adequate for our purposes to take for the quasicrystals a straightforward concatenation of one-dimensional ones in pairwise orthogonal directions.

\subsection{Computational aspects}
First let us introduce two specific examples of
one-dimensional quasicrystals $\Lambda_{\tau}(\Omega)$.
\begin{example}
Take $\tau=\frac{1+\sqrt{5}}{2}$, $\tau'=\frac{1-\sqrt{5}}{2}$, the
two roots of the quadratic equation $x^2=x+1$. Then $\tau+\tau'=1$,
$\tau\tau'=-1$. Choosing $\Omega=[0,d)$, $d=\frac{\tau}{2}$ the
conditions of Theorem~2 are  satisfied:
$\tau'\approx -0.62\in(-1,0)$ and $\tau\approx 1.62>0$.

Due to Theorem~2,
the distances between adjacent points of the quasicrystal $\Lambda_{\tau}([0,d))$ are
$\Delta_1=\tau$, $\Delta_2=1+\tau$ and $\Delta_3=1+2\tau$.

As the seed point of the quasicrystal $\Lambda_{\tau}([0,d))$,
we can choose any point $x=a+b\tau$, such that $x'=a+b\tau'\in [0,d)$.
For our example we put $x=0$.

Unlike Definition 1, which is not constructive,
here we have the information for fast generation of quasicrystal points.
We can move right or left from the seed point by adding or subtracting
one-by-one of the distances $\Delta_1$, $\Delta_2$ or $\Delta_3$.

Suppose we have already established, that $x$ is a point of $\Lambda_{\tau}([0,d))$, i.e. $x'\in [0,d)$.
The point adjacent to $x$ from the right is one of the three
\begin{gather*}
x+\tau,\quad x+1+\tau\quad \text{or}\quad x+1+2\tau.
\end{gather*}

In order to decide which one is the case, we verify one-by-one the inclusions
$x+\tau'\in [0,d)$, $x+1+\tau'\in [0,d)$ or $x+1+2\tau'\in [0,d)$.
The first confirmed inclusion determines the new quasicrystal point.

In such a way the following two subsets of the quasicrystal
$\Lambda_{\tau}([0,d))$ were obtained
\begin{gather*}
\{\ldots, -1\!-\!2\tau,\ -\tau,\ 0,\ 1\!+\!\tau,\ 2\!+\!2\tau,\ 2\!+\!3\tau,\ 3\!+\!4\tau, \ldots\} \leftrightarrow
\\
\qquad\qquad\qquad
\leftrightarrow\{\ldots,\ \Delta_2,\ \Delta_1,\ \Delta_2,\ \Delta_2,\ \Delta_1,\ \Delta_2,\ \ldots\};
\\
\!\{\ldots,-\tau,\ 0,\ 1+\tau,\ 2+3\tau,\ 3+4\tau,\ldots\}\!\leftrightarrow
\\
\qquad\qquad\qquad
\leftrightarrow\{\ldots,\Delta_1,\ \Delta_2,\ \Delta_3,\ \Delta_2,\ldots\}.
\end{gather*}

Here the second line of each example contains the corresponding sequence of distances between quasicrystal points.
The two quasicrystals differ by the sequence in which the three distances
$\Delta_1$, $\Delta_2$ and $\Delta_3$ were tried in the construction.

The first case is was shown as the second line on Fig.~\ref{fig_shema}.
\end{example}

\begin{example}
Next quasicrystal $\Lambda_{\tau}(\Omega)$ is built using the irrationalities $\tau=1+\sqrt{2}$, $\tau'=1-\sqrt{2}$,
roots of the quadratic equation $x^2=2x+1$.

The conditions of Theorem~2 are again satisfied:\linebreak
$\tau'\approx-0.42\in(-1,0)$ and $\tau\approx 2.42>0$.
Putting $\Omega=[0,1)$,
there are only two distances $\Delta_1=\tau$ and $\Delta_2=1+\tau$ between adjacent points of the quasicrystal
$\Lambda_{\tau}([0,1))$.

Using the zero point as the seed point, we obtain the cut and project set
\begin{gather*}
\{\ldots,\!-\!1\!-\!3\tau,
\!-\!2\tau,\!-\!\tau,0,1\!+\!\tau,1\!+\!2\tau,2\!+\!3\tau,
2\!+\!4\tau,3\!+\!5\tau,\ldots\}
\\
\qquad%
\leftrightarrow\{\ldots,\ \Delta_2,\ \Delta_1,\ \Delta_1,\
\Delta_2,\ \Delta_1,\ \Delta_2,\ \Delta_1,\ \Delta_2,\ \ldots\}.
\end{gather*}
\end{example}

It may happen that some point $x'$ is arbitrary close to an end point of $\Omega$.
Then it would be a time demanding computational task to decide whether $x'\in \Omega$ or $x'\notin \Omega$.
Such a difficulty is simply avoided by disqualifying from our consideration any point $x'$
which comes closer to the boundary of $\Omega$ than the distance $\varepsilon$ agreed in advance.

The real numbers of the form $a+b\tau$ can be understood as given by two integer components $(a,b)$.
Since our operations are only addition or subtraction, the arithmetics of such numbers is elementary introduced.
Consequently all transformations are performed with absolute precision.

The transformations between the original and encrypted data go through the several stages.
At each stage we need only the~previous point.
Therefore we have only minimal requirements to RAM.

\section{Application to image coding}\label{sec_image_coding}
In this section fragments of quasicrystal point sets are used for
encryption/decryption of 2-dimensional digital data.

Assuming that the data is given on the rectangular grid containing final number $n_1$, $n_2$ of points in each direction,
we deal with final fragments $\Lambda^{(i)}(\Omega_i)$, $i=1,2$
of the quasicrystal $\Lambda(\Omega)=\Lambda^{(1)}(\Omega_1)\otimes
\Lambda^{(2)}(\Omega_2)$.
Number of points in each fragment  is given by digital data size $n_i=|\Lambda^{(i)}(\Omega_i)|$.

{\looseness=-1
Each $\Omega_i$ is an interval on real axis in
mutually or\-tho\-go\-nal directions.
They can be chosen for both fragments independently.

}

Our main tool is the one-to-one map (star map)
between the finite point sets $\Lambda_{i}(\Omega_i)$ and
its image in $\Omega_i$
\begin{gather*}\label{themap}
\Lambda_{i}(\Omega_i)\quad\longleftrightarrow\quad
     \Omega_i,\qquad i=1,2\,.
\end{gather*}

Note, that roles of $\Lambda_{i}(\Omega_i)$ and
its image in $\Omega_i$ can be interchanged, because the fragments
$\Lambda_{i}(\Omega_i)$ are finite sets of~points.

Fast generation of $\Lambda_{i}(\Omega_i)$ was described in previous section.
It requires that one is given the values of
$n_1$, $n_2$, $\tau_i$, $\tau_i'$ and $\Omega_i$,
and two seed points of $\Lambda_{i}(\Omega_i)$, $i=1,2$.

Suppose one is given a bitmap picture $P(L)$,
sampled on the points $(x,y)$ of a fragment $L$ of a rectangular lattice.

Then we treat independently each coordinate
of the data using one fragment of the quasicrystal as illustrated in Fig.~\ref{fig_shema}.

\subsection{Basic encryption algorithm}\label{subsec_main_algo}
\begin{itemize}
\item
Construction of two different 1-dimensional quasicrystals
$\Lambda^{(1)}$ and $\Lambda^{(2)}$ of sizes $n_1$ and $n_2$ appropriate for the given data $P(L)$.
(Each fragment, as well as its acceptance window--interval can be situated anywhere on the corresponding real axes.)
\bigskip

\item
Matching one-by-one the sequence of coordinates $x$ with points of $\Lambda^{(1)}$, similarly matching the
coordinates $y$ with points of
$\Lambda^{(2)}$.
(In this way $x$ becomes point $x_1+x_2\tau$ of the $\Lambda^{(1)}$ and $y$ becomes point $y_1+y_2\tau$ of the $\Lambda^{(2)}$, where $x_1$, $x_2$, $y_1$, $y_2$ are integers.)
\bigskip

\item
The main step of the encoding is the star map of
$\Lambda^{(1)}\otimes\Lambda^{(2)}$
into the corresponding window $\Omega_1\otimes\Omega_2$
\begin{gather*}
\begin{tabular}{ccc}
$\Lambda^{(1)}(\Omega_1)\longrightarrow\Omega_1$,&\qquad&
$\Lambda^{(2)}(\Omega_2)\longrightarrow\Omega_2$;
\\
$x_1+x_2\tau_1 \mapsto x_1+x_2\tau_1'$,&&
$y_1+y_2\tau_2 \mapsto y_1+y_2\tau_2'$.
\end{tabular}
\end{gather*}
(The sequence of the original data coordinates is dra\-ma\-ti\-cal\-ly changed due to the discontinuity
of the star map.)
\smallskip

\item
It remains to map the points of encrypted quasicrystal fragments, which are now in  $\Omega_1$ and $\Omega_2$,
to a sequence of integers along $x$ and $y$ directions.
\end{itemize}


The present algorithm permutes the data points
without changing the value of the data at each point.
It is straightforward to consider the possible values of the data
function as another discrete point set and to apply the same algorithm to it.
Effectively we are thus dealing with a three-dimensional problem,
where the data function takes only two values at 3-dimensional points.

All the maps we have used in the encryption algorithm are one-to-one,
therefore they can be used in the inverse order for the decryption.
For the same reason, nothing prevent us from iterative using our algorithm.

\begin{figure}[ht]
\centerline{\includegraphics[scale=1]{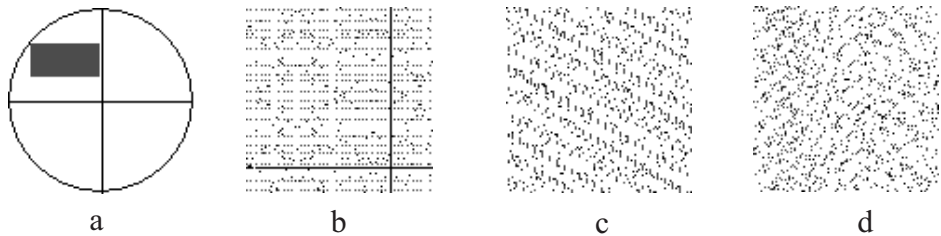}}
\caption{Application of the algorithms proposed in Section~\ref{sec_image_coding}:
a -- initial image; b -- main algorithm;  c -- second modification; d -- first modification}\label{fig0}
\end{figure}

In certain types of data the encryption algorithm leaves
residue of the orthogonal directions, see Fig.~\ref{fig0}b.
It can be avoided in a many different ways. One of them was mentioned above,
when the problem was viewed as a three-dimensional one.
Another possibility is illustrated on Fig.~\ref{fig0}c.
The two-dimensional data is taken as an ordered one-dimensional sequence (modification 2).
The particularly fast possibility is to introduce the additional cyclic permutations
of the points (modification 1), see Fig.~\ref{fig0}d.

The computation time of our algorithms is compared in Table~1.
Computations were carried out by Pentium M processor 1.70 GHz with 1 Gb RAM.
As the test image we took the 512x512-pixel 8 bit grayscale picture Lenna (see Figs.~\ref{fig3} and~\ref{fig4}).
But the second modification was tested on 200x200-pixel 8 bit grayscale picture Lenna (Fig.~5).
Note, that the program does not cut the images into blocks for processing.

\begin{center}
Table~1. Program working time.
\smallskip

\begin{tabular}{|c|c|c|c|}
\hline
&&&\\[-5 pt]
& Main algorithm &Modification 1 &Modification 2\\[-7 pt]
&&&\\
\hline
&&&\\[-5 pt]
time& 3 seconds & 6 seconds & 192 seconds\\[-7 pt]
&&&\\
\hline
\end{tabular}
\end{center}

Program realizing these algorithms can be freely downloaded from
\textit{http://132.204.90.210/maryna/} and listing of the main
coding algorithm is adduced in Appendix~\ref{app_program}.

\section{Conclusions}
In the paper one of the simplest versions of the quasicrystal-based symmetric cipher is described.
Even in this form it appears adequate for many applications.
Its complicated key is described below.
It is based on strongly aperiodic sets, includes everywhere discontinuous `star map', and it is absolutely stable,
since it operates with integer numbers only.
Similarly, damaging a few data points does not affect the rest of the encryption/decryption process.

Furthermore coordinates of the data can be divided
into several non-overlapping  intervals, then contents of all intervals could be encrypted in parallel.

\subsection{Cryptovariables}
Application of the algorithm requires that the following
variables (cryptovariables of the private key) are fixed
\begin{gather*}
\tau\_\tau'\_a\_b\_c\_d\_sml\_X\_\varepsilon\_I,
\quad \text{where}
\end{gather*}
$\tau$ and $\tau'$ are irrational numbers, solutions of
one of equations of the two infinite series of the quadratic equations \mbox{$x^2=mx+1$}, $m=1,2,\ldots$
or $x^2=mx-1$, $m=3,4,\ldots$;
\smallskip

\noindent
$a$ and $b$ are integers, fixing a seed point of the quasicrystal;
\smallskip

\noindent
$c$ and $d$  are arbitrary real numbers, fixing the position of the acceptance window $\Omega$;
\smallskip

\noindent
$X$ takes values \{`+', `-'\}, defining direction of quasicrystal construction relative to the seed point;
\smallskip

\noindent
$sml$ sequence of distances of adjacent points during quasicrystal generation;
\smallskip

\noindent
$\varepsilon$ is the maximal distance from the end points of $\Omega$ at which the star map points are still considered;
\smallskip

\noindent
$I$ is the positive integer number specifying number of iterations of the certain quasicrystal cipher.
\bigskip

More elaborate encryption system can be built by exploiting
properties of 1 or 2-dimensional quasicrystals.
Using the present algorithm in several  iterations,
one may choose new quasicrystal for each iteration,
 or new position of the quasicrystal fragment,
 or different size, position of the acceptance window, etc.

Rich scaling properties of the quasicrystal could be involved into encryption
scheme~\cite{Gazeau-Masakova-Pelantova2006}.

Genuinely two-dimensional quasicrystals can be used for the encryption,
some examples of the possibilities are found in~\cite{patera-nesterenko}.
The number of distinct quasicrystals of this kind is unlimited.

One can easily build one-dimensional quasicrystal by projecting cuts of
n-dimensional latices on the direction of the quasicrystal.

Intriguing appears to be the possibility to read the initial lattice data
as an ordered stream of points and map them into single one-dimensional quasicrystal,
perform the star map and reorder transform points into encrypted data.
There are also other applications of the quasicrystals e.g. for generation of
random numbers~\cite{Guimond-Patera-Patera2003}, even some cryptographic ones~\cite{patent}.

\section*{Acknowledgment}
One of us (M.N.) is grateful for the hospitality extended to her at the Center de recherche math\'ematique,
Universit\'e de Montr\'eal during the work on this project.

We are grateful for partial support of the work by the Natural
Science and Engineering Research Council of Canada, MITACS,
Lockheet Martin of Canada and MIND Research Institute of California.

\appendix

\section{Program listing}\label{app_program}
Realization of the main algorithm on Delphi 6.0:

unit Unit1;

interface

uses

\ Windows, Messages, SysUtils, Variants, Classes, Graphics,
Controls, Forms,

\ Dialogs, StdCtrls;

type

\ \ \ TNode = Record

\ \ \ \ \ \ \ a \ \ \ \ \ \ : integer;

\ \ \ \ \ \ \ b \ \ \ \ \ \ : integer;

\ \ \ \ \ \ \ rs, rb \ : real;

\ \ \ \ \ \ \ ns, nb \ : integer;

\ \ \ end;

\ \ \ TSequence = Record

\ \ \ \ \ \ \ n \ \ \ \ \ \ : integer;

\ \ \ \ \ \ \ Nodes \ \ : array of TNode;

\ \ \ \ \ \ \ a,b \ \ \ \ : integer;

\ \ \ \ \ \ \ s \ \ \ \ \ \ : string;

\ \ \ end;

\ \ \ TForm1 = class(TForm)

\ \ \ Button1 \ \ \ \ \ \ \ \ : TButton;

\ \ \ Button2: TButton;

\ \ \ procedure Button1Click(Sender: TObject);

\ \ \ procedure FormCreate(Sender: TObject);

\ \ \ procedure Button2Click(Sender: TObject);

\ private

\ \ \ { Private declarations }

\ public

\ \ \ { Public declarations }

\ end;

var

\ Form1 \ \ \ \ \ \ \ \ : TForm1;

\ RootS, RootB \ : real;

implementation

\text{\{\$R *.dfm\}}

function CalcSmall (a, b: real) : real;

begin

\ \ \ CalcSmall := a + b * RootS;

end;

function CalcBig (a, b: real) : real;

begin

\ \ \ {a := a + 1;

\ \ \ b := - b;

\ \ \ CalcBig := CalcSmall (a, b);}

\ \ \ CalcBig := a + b * RootB;

end;

function CheckShort (a, b: integer; var Node : Tnode; dir: integer)
: boolean;

var tempB : real;

\ \ \ res : boolean;

begin

\ \ \ a := a - dir;

\ \ \ b := b + dir;

\ \ \ tempB := CalcBig (a,b);

\ \ \ res := ((tempB $<$ 0) and (tempB $<$ 2));

\ \ \ If res then begin

\ \ \ \ \ \ \ node.a := a;

\ \ \ \ \ \ \ node.b := b;

\ \ \ \ \ \ \ node.rb := tempB;

\ \ \ end;

\ \ \ CheckShort := res;

end; //CheckShort (a, b: integer; var node : Tnode; dir: integer) :
boolean;

function CheckMedium (a, b: integer; var node : Tnode; dir: integer)
: boolean;

var tempB : real;

\ \ \ res : boolean;

begin

\ \ \ a := a + dir;

\ \ \ tempB := CalcBig (a,b);

\ \ \ res := ((tempB $>$ 0) and (tempB $<$ 2));

\ \ \ If res then begin

\ \ \ \ \ \ \ node.a := a;

\ \ \ \ \ \ \ node.b := b;

\ \ \ \ \ \ \ node.rb := tempB;

\ \ \ end;

\ \ \ CheckMedium := res;

end; //CheckMedium (a, b: integer; var node : Tnode; dir: integer) :
boolean;

function CheckLong (a, b: integer; var node : Tnode; dir: integer) :
boolean;

var tempB : real;

\ \ \ res : boolean;

begin

\ \ \ b := b + dir;

\ \ \ tempB := CalcBig (a,b);

\ \ \ res := ((tempB $>$ 0) and (tempB $<$ 2));

\ \ \ If res then begin

\ \ \ \ \ \ \ node.a := a;

\ \ \ \ \ \ \ node.b := b;

\ \ \ \ \ \ \ node.rb := tempB;

\ \ \ end;

\ \ \ CheckLong := res;

end; //CheckLong (a, b: integer; var node : Tnode; dir: integer) :
boolean;

function FillSequence (var seq : TSequence; a, b, n, dir: integer):
string;

var res \ \ \ \ \ \ \ \ \ \ \ \ : string;

\ \ \ aPrev, bPrev \ \ \ : integer;

\ \ \ i \ \ \ \ \ \ \ \ \ \ \ \ \ \ : integer;

begin

\ \ \ res := '';

\ \ \ seq.n := n;

\ \ \ seq.a := a;

\ \ \ seq.b := b;

\ \ \ SetLength (seq.Nodes, n);

\ \ \ seq.Nodes[0].a := a;

\ \ \ seq.Nodes[0].b := b;

\ \ \ seq.Nodes[0].rb := CalcBig (a, b);

\ \ \ seq.Nodes[0].rs := CalcSmall (a, b);

\ \ \ seq.Nodes[0].ns := 0;

\ \ \ for i := 1 to n - 1 do begin

\ \ \ \ \ \ \ aPrev := seq.Nodes [i-1].a;

\ \ \ \ \ \ \ bPrev := seq.Nodes [i-1].b;

\ \ \ \ \ \ \ seq.Nodes [i].ns := i;

\ \ \ \ \ \ \ if dir = 1 then begin

\ \ \ \ \ \ \ \ \ \ \ if CheckShort (aPrev, bPrev, seq.Nodes [i],
dir) then

\ \ \ \ \ \ \ \ \ \ \ \ \ \ \ res := res + 's'

\ \ \ \ \ \ \ \ \ \ \ else if CheckMedium (aPrev, bPrev, seq.Nodes
[i], dir) then

\ \ \ \ \ \ \ \ \ \ \ \ \ \ \ res := res + 'm'

\ \ \ \ \ \ \ \ \ \ \ else if CheckLong (aPrev, bPrev, seq.Nodes
[i], dir) then

\ \ \ \ \ \ \ \ \ \ \ \ \ \ \ res := res + 'l'

\ \ \ \ \ \ \ end else if CheckLong (aPrev, bPrev, seq.Nodes [i],
dir) then

\ \ \ \ \ \ \ \ \ \ \ \ \ \ \ res := res + 'l'

\ \ \ \ \ \ \ \ \ \ \ else if CheckMedium (aPrev, bPrev, seq.Nodes
[i], dir) then

\ \ \ \ \ \ \ \ \ \ \ \ \ \ \ res := res + 'm'

\ \ \ \ \ \ \ \ \ \ \ else if CheckShort (aPrev, bPrev, seq.Nodes
[i], dir) then

\ \ \ \ \ \ \ \ \ \ \ \ \ \ \ res := res + 's';

\ \ \ \ \ \ \ seq.Nodes[i].rs := CalcSmall (seq.Nodes[i].a,
seq.Nodes[i].b);

\ \ \ end;

\ \ \ seq.s := res;

\ \ \ FillSequence := res;

end; //FillSequence (var seq : TSequence; a, b, n, dir: integer):
string;

Procedure SaveSequence (seq : TSequence; fname : string);

var f : TextFile;

var i : integer;

begin

\ \ \ AssignFile (f, fname);

\ \ \ rewrite (f);

\ \ \ writeln (f, seq.n,' : ', seq.s);

\ \ \ for i:= 0 to seq.n - 1 do

\ \ \ \ \ \ \ writeln (f, seq.Nodes[i].ns, ', ', seq.Nodes[i].nb, ',
', seq.Nodes[i].a, ', ', seq.Nodes[i].b, ', ', seq.Nodes[i].rs, ',
', \ seq.Nodes[i].rb);

\ \ \ Flush (f);

\ \ \ CloseFile (f);

end; //SaveSequence (seq : TSequence; fname : string);

Procedure SortSequence (var seq : TSequence; dir: integer);

var node \ \ \ : TNode;

\ \ \ i,j \ \ \ \ : integer;

\ \ \ flag \ \ \ : integer;

begin

\ \ \ for i := seq.n-2 downto 0 do

\ \ \ \ \ for j:= 0 to i do begin

\ \ \ \ \ \ \ flag := 0;

\ \ \ \ \ \ \ if dir = 1 then begin

\ \ \ \ \ \ \ \ \ \ \ if seq.Nodes[j].rb $>$ seq.Nodes[j + 1].rb
then

\ \ \ \ \ \ \ \ \ \ \ \ \ \ \ flag := 1;

\ \ \ \ \ \ \ end else

\ \ \ \ \ \ \ \ \ \ \ if seq.Nodes[j].rs $>$ seq.Nodes[j + 1].rs
then

\ \ \ \ \ \ \ \ \ \ \ \ \ \ \ flag := 1;

\ \ \ \ \ \ \ if flag = 1 then begin

\ \ \ \ \ \ \ \ \ \ \ node := seq.Nodes[j + 1];

\ \ \ \ \ \ \ \ \ \ \ seq.Nodes[j + 1] := seq.Nodes[j];

\ \ \ \ \ \ \ \ \ \ \ seq.Nodes[j] := node;

\ \ \ \ \ \ \ end;

\ \ \ \ \ end;

\ \ \ if dir = 1 then

\ \ \ \ \ \ \ for i := 0 to seq.n do

\ \ \ \ \ \ \ \ \ \ seq.Nodes[i].nb := i;

end; //SorteSequence (var seq : TSequence; dir: integer)

Procedure CodeImg (dir: integer);

var fname1, fname2 \ \ \ \ \ : string;

\ \ \ Bitmap1, Bitmap2 \ \ \ : TBitmap;

\ \ \ SeqX, SeqY \ \ \ \ \ \ \ \ \ : TSequence;

\ \ \ i,j \ \ \ \ \ \ \ \ \ \ \ \ \ \ \ \ : integer;

\ \ \ CurWidth, CurHeight : integer;

\ \ \ strX, strY \ \ \ \ \ \ \ \ \ : string;

begin

\ \ \ if dir = 1 then begin

\ \ \ \ \ \ \ fname1 := 'in.bmp';

\ \ \ \ \ \ \ fname2 := 'out.bmp';

\ \ \ end else begin

\ \ \ \ \ \ \ fname1 := 'out.bmp';

\ \ \ \ \ \ \ fname2 := 'back.bmp';

\ \ \ end;

\ \ \ Bitmap1 := TBitmap.Create;

\ \ \ Bitmap1.LoadFromFile (fname1);

\ \ \ Bitmap2 := TBitmap.Create;

\ \ \ CurWidth := Bitmap1.Width;

\ \ \ CurHeight := Bitmap1.Height;

\ \ \ Bitmap2.Width := CurWidth;

\ \ \ Bitmap2.Height := CurHeight;

\ \ \ StrX := FillSequence (SeqX, 1, 0, CurWidth, 1);

\ \ \ StrY := FillSequence (SeqY, 1, 0, CurHeight, 1);

\ \ \ SortSequence (SeqX, 1);

\ \ \ SortSequence (SeqY, 1);

\ \ \ for i:= 0 to CurWidth - 1 do

\ \ \ \ \ \ \ for j:= 0 to CurHeight - 1 do

\ \ \ \ \ \ \ \ \ \ \ if dir = 1 then

\ \ \ \ \ \ \ \ \ \ \ \ \ \ \ Bitmap2.Canvas.Pixels [i,j] :=
Bitmap1.Canvas.Pixels [SeqX.Nodes [i].ns, SeqY.Nodes [j].ns]

\ \ \ \ \ \ \ \ \ \ \ else

\ \ \ \ \ \ \ \ \ \ \ \ \ \ \ Bitmap2.Canvas.Pixels [SeqX.Nodes
[i].ns,SeqY.Nodes [j].ns] := Bitmap1.Canvas.Pixels [i,j];

\ \ \ Bitmap2.SaveToFile (fname2);

\ \ \ Bitmap1.Free;

\ \ \ Bitmap2.Free;

\ \ \ if dir = 1 then begin

\ \ \ \ \ \ \ SortSequence (SeqX, -1);

\ \ \ \ \ \ \ SortSequence (SeqY, -1);

\ \ \ \ \ \ \ SaveSequence (SeqX, 'x.txt');

\ \ \ \ \ \ \ SaveSequence (SeqY, 'y.txt');

\ \ \ end;

end; //CodeImg (dir: integer)

procedure TForm1.Button1Click(Sender: TObject);

begin

\ \ \ Button1.Enabled := False;

\ \ \ Button2.Enabled := False;

\ \ \ CodeImg (1);

\ \ \ Button1.Enabled := True;

\ \ \ Button2.Enabled := True;

end;

procedure TForm1.FormCreate(Sender: TObject);

begin

\ \ \ RootS := (1 + sqrt (5)) / 2;

\ \ \ RootB := (1 - sqrt (5)) / 2;

end;

procedure TForm1.Button2Click(Sender: TObject);

begin

\ \ \ Button1.Enabled := False;

\ \ \ Button2.Enabled := False;

\ \ \ CodeImg (-1);

\ \ \ Button1.Enabled := True;

\ \ \ Button2.Enabled := True;

end;

end.

\section{Application examples}\label{app_pictures}
In this section we will show several examples of the application  of
the main algorithm, its modification 1, and modification 2, as well
as its iterations several times. Let us emphasize that the data was
encrypted as a single block. Computing time is ranging between two
seconds (one iteration) and nine seconds (five iterations) for the
main algorithm.

\begin{figure}[ht]
\centerline{\includegraphics[scale=1]{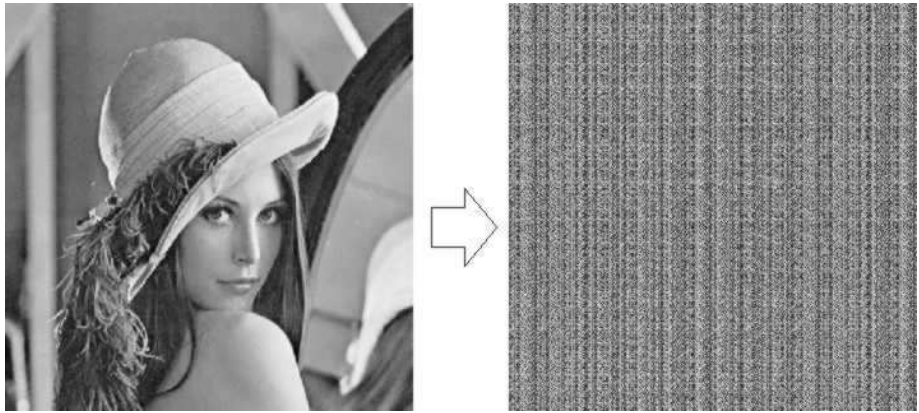}}
\caption{Main algorithm coding without iterations}\label{fig3}
\end{figure}

\begin{figure}[ht]
\centerline{\includegraphics[scale=1]{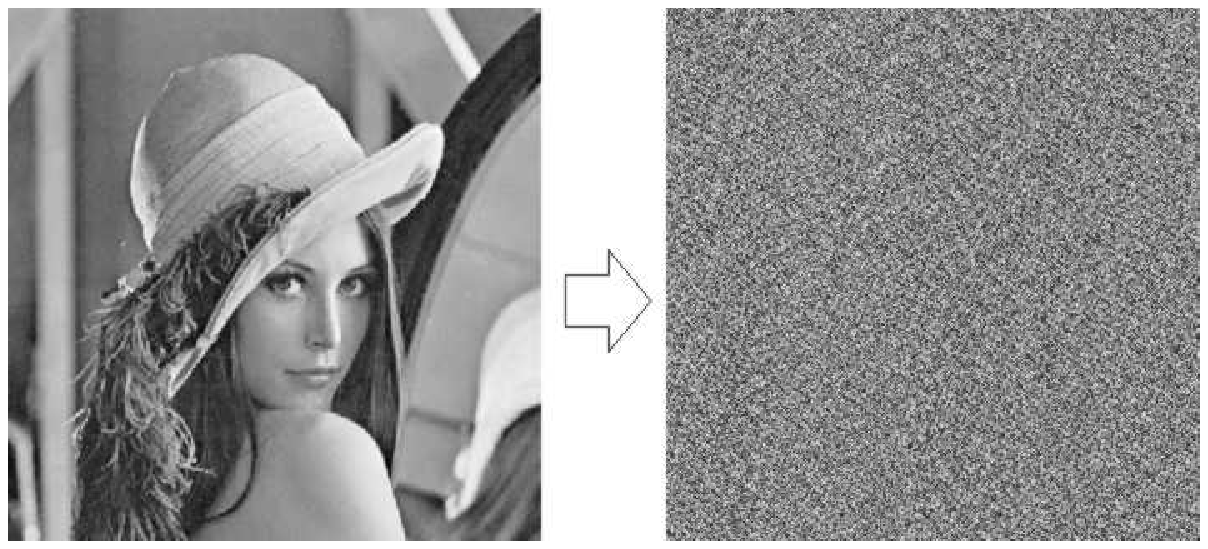}}
\caption{Modification 1 of the algorithm without
iterations}\label{fig4}
\end{figure}

\begin{figure}[ht]
\centerline{\includegraphics[scale=1]{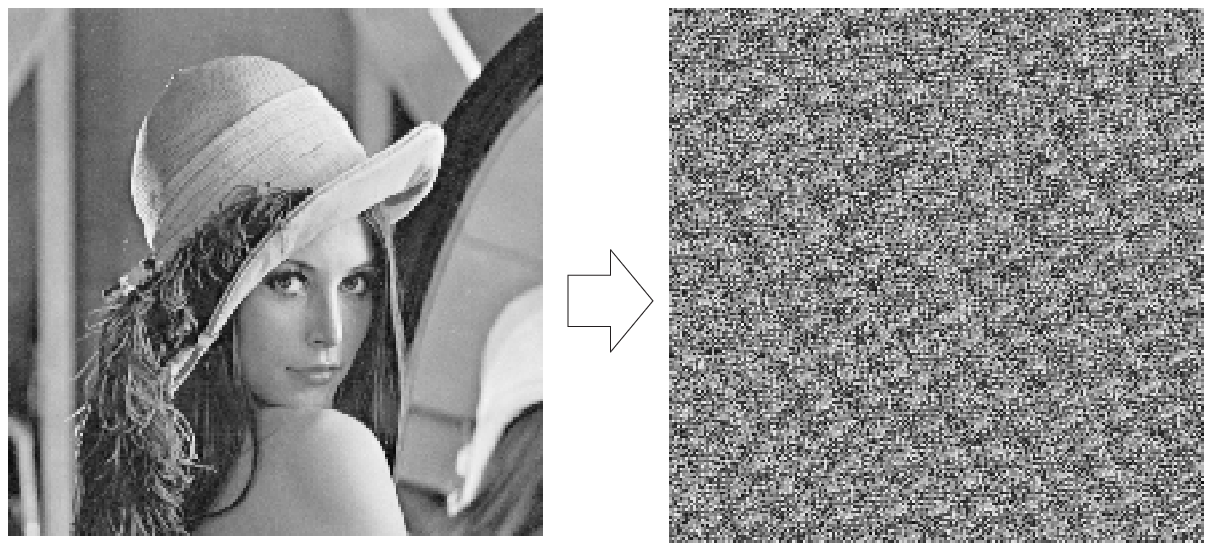}}
\caption{Modification 2 of the algorithm without iterations}\label{fig5}
\end{figure}

\begin{figure}[ht]
\centerline{\includegraphics[scale=1]{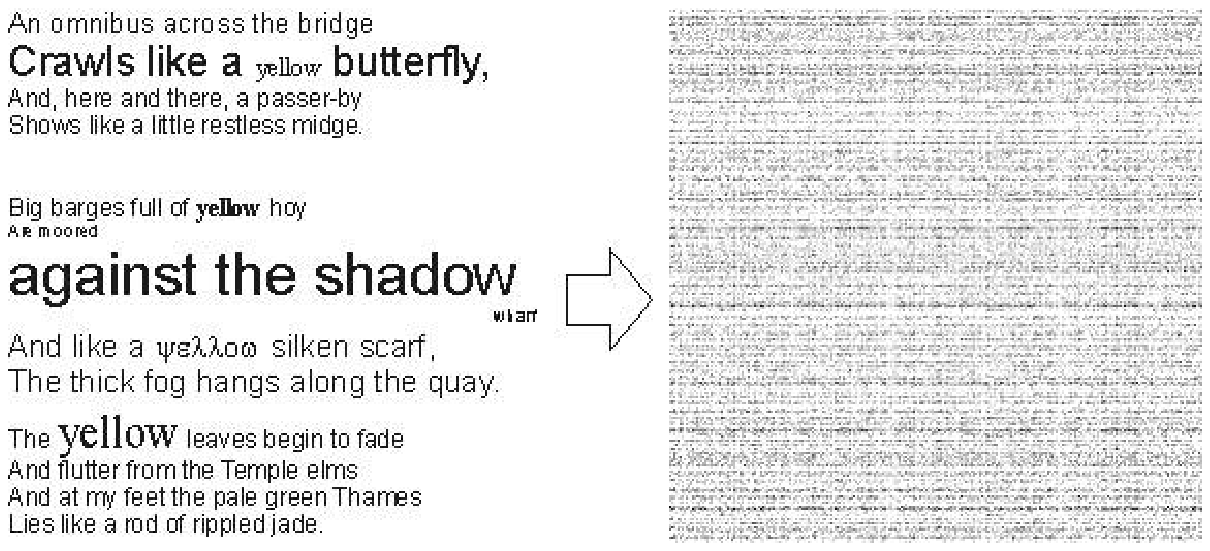}}
\caption{Main algorithm coding without iterations}\label{fig9}
\end{figure}

\begin{figure}[ht]
\centerline{\includegraphics[scale=1]{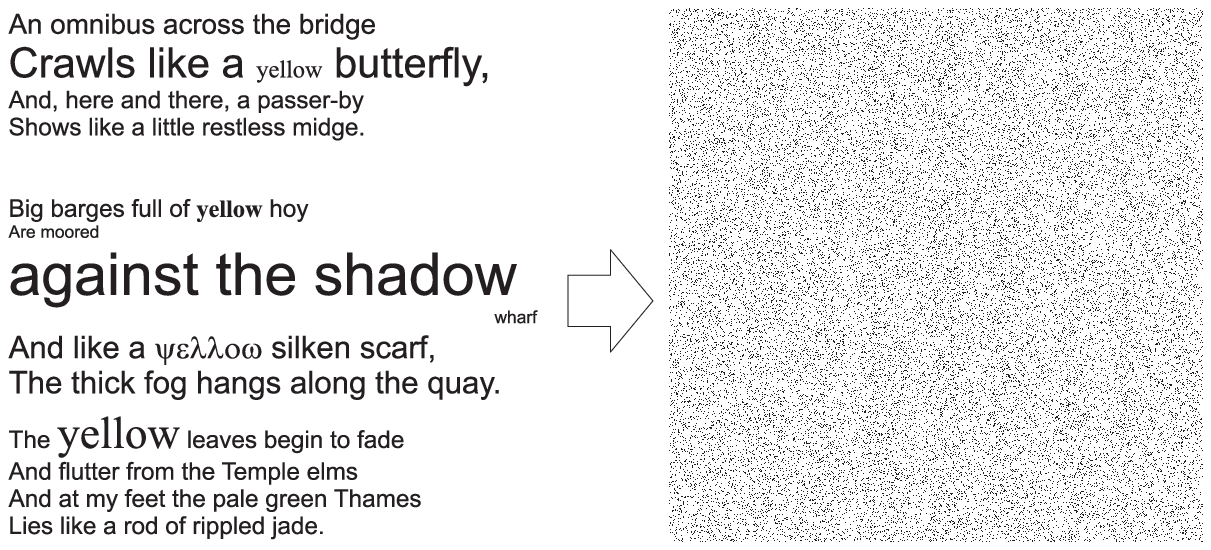}}
\caption{Modification 1 of the algorithm without iterations}\label{fig10}
\end{figure}
\end{document}